\documentclass[11pt]{iopart}
\usepackage[english]{babel}
\usepackage{graphicx}
\begin{document}
\title{Astrophysical Sources of Stochastic Gravitational-Wave Background}
\author{T. Regimbau$^1$ and V. Mandic$^2$}
\address{$^1$ UMR ARTEMIS, CNRS, University of Nice Sophia-Antipolis, Observatoire de la C\^ote d'Azur, BP 429
06304 Nice (France)\\ $^2$ School of Physics and Astronomy, University of Minnesota-Twin Cities (USA)}
\ead{regimbau@oca.eu, mandic@physics.umn.edu}

\begin{abstract}
We review the spectral properties of stochastic backgrounds of astrophysical origin and discuss how they may differ from the primordial contribution by their statistical properties.
We show that stochastic searches with the next generation of terrestrial interferometers could put interesting constrains on the physical properties of astrophysical populations, such as the ellipticity and magnetic field of magnetars, or the coalescence rate of compact binaries.

\end{abstract}

\section{Introduction}

According to various cosmological scenarios, we are
bathed in a stochastic background of gravitational waves, memory of
the first instant of the Universe, up to the limits of the Plank
era and the Big Bang \cite{gri01,mag00}.
In addition to the cosmological background (CGB), an astrophysical
contribution (AGB) may have resulted from the superposition of a large
number of unresolved sources since the beginning of stellar
activity, which can be either short lived burst sources, such as core
collapses to neutron stars \cite{bla96,cow01,how04,buo04} or black holes \cite{fer99a,dara00,dara02a,dara02b,dara04},  phase transition or oscillation
modes in young neutron stars \cite{owe98,fer99b,sig06}, the final stage of compact binary mergers \cite{reg06b,reg07}, or periodic long lived sources,
typically pulsars \cite{reg01a,reg06a}, the early inspiral phase of compact
binaries \cite{ign01,sch01,far02,coo04} or captures by supermassive black holes \cite{bar04,schn06},
whose frequency is expected to evolve very slowly compared to the observation time.

The astrophysical contribution is important for at least two reasons:
on one hand, it carries information about the star formation history, the mass range of neutron star or black hole progenitors, the statistical properties of populations of compact objects like the
ellipticity or the magnetic field of neutron stars, or the rate of compact binary mergers.
On the other hand, it can be a foreground for the cosmological background, and it has to be modeled accurately to define the best frequency windows where to search for the cosmological background.
The rest of this article will be organized as follows: in Sec. 2, we review the spectral properties of the AGBs; in Sec. 3, we discuss how their statistical properties may differ from those of the cosmological background; in Sec. 4 we present two simple models, the stochastic background from magnetars and double neutron star coalescences, susceptible of producing a continuous stochastic background in the frequency band of terrestrial interferometers; in Sec. 5 we discuss the constraints one can expect to put on these models and on the source populations  with the next generations of detectors. Finally, in Sec. 6 we summarize the results and present our conclusions.

\section{The spectral properties}

The spectrum of the gravitational stochastic background is usually
characterized by the dimensionless parameter:

\begin{equation}
\Omega_{gw}(\nu_o)=\frac{1}{\rho_c}\frac{d\rho_{gw}}{d\ln \nu_o}
\end{equation}
where $\rho_{gw}$ is the gravitational energy density, $\nu_o$ the
frequency in the observer frame and $\rho_c=\frac{3H_0^2}{8 \pi G}$
the critical energy density needed to close the Universe
today.
For a stochastic background of astrophysical origin, the energy density is given by:
\begin{equation}
\Omega_{gw}(\nu_o)=\frac{1}{\rho_c c^3} \nu_o F_{\nu_o}(\nu_o)
\end{equation}
where the integrated flux at the observed frequency $\nu_o$ is defined as:
\begin{equation}
F_{\nu_o}(\nu_o)=\int f_{\nu_o}(\nu_o,z) \frac{dR(z)}{dz}dz
\label{eq-flux}
\end{equation}
The spectral properties of a single source located at z are given by
the fluence:
\begin{equation}
f_{\nu_o}(\nu_o,z)=\frac{1}{4 \pi
  r(z)^2} \frac{dE_{gw}}{d \nu}(\nu_o(1+z))
\end{equation}
where $r$ the proper
distance, which depends on the adopted cosmology, $\frac{dE_{gw}}{d\nu}$ the gravitational spectral energy emitted by a single source
and $\nu=\nu_o(1+z) $ the frequency in the source frame.

When the gravitational emission occurs shortly after the birth of the
progenitor, the event rate
can be associated with the progenitor formation rate:
\begin{equation}
\frac{dR(z)}{dz}=\lambda \frac{R_*(z)}{1+z} \frac{dV}{dz}(z)
\label{eq-rate}
\end{equation}
where $R_*(z)$ is the cosmic star formation rate per comoving volume
(SFR) expressed in M$_\odot$ Mpc$^{-3}$ yr$^{-1}$ and $\lambda$ the mass fraction
converted into the progenitors, assumed to be the same at all redshifts.
The $(1+z)$ term in the denominator corrects the cosmic star formation
rate by the time dilatation due to the cosmic expansion.
In our calculations, we consider the 737 cosmology \cite{rao06}, with $\Omega_m=0.3$,
$\Omega_{\Lambda}=0.7$ and Hubble parameter $H_0=70$ km s$^{-1}$
Mpc$^{-1}$ (or equivalently $h_0=0.7$), corresponding to the so-called concordant
model derived from observations of distant type Ia supernovae
\cite{per99} and the power spectra of the cosmic microwave background
fluctuations \cite{spe03}.
We use the recent SFR
of \cite{hop06} derived from new measurements of the galaxy luminosity function in the
UV (SDSS, GALEX, COMBO17) and FIR wavelengths (Spitzer Space
Telescope), which allowed to refine the previous models up to redshift $z \sim 6$, with very tight constraints
at redshifts $z<1$ (and quite accurate up to $z \sim 2$), and normalized by the Super Kamiokande limit on the electron antineutrino flux from past core-collapse supernovas.
A parametric fit is given by \cite{hop06}:
\begin{equation}
R_*(z)= h_0 \frac{0.017 + 0.13z}{1 + (z/3.3)^{5.3}}
\label{eq-sfr}
\end{equation}
for an initial mass function (IMF) of the form (modified Salpeter A):
\begin{equation}
\xi(m) \: \propto
\left\lbrace
\begin{array}{ll}
(\frac{m}{m_0})^{-1.5}    &   \hbox{  for } 0.1<m<m_0 \\
(\frac{m}{m_0})^{-2.35} &   \hbox{  for } m_0<m<100 \\
\end{array}
\right.
\label{eq-imf}
\end{equation}
with a turnover below $m_0=0.5$ M$_{\odot}$ and normalized within the mass interval $0.1 - 100$ M$_{\odot}$ such as $\int m\xi(m)dm$ = 1.
The element of comoving volume in eq.~\ref{eq-rate} is given by:
\begin{equation}
\frac{dV}{dz}(z)=4 \pi \frac{c}{H_0} \frac{r(z)^2}{E(\Omega,z)}
\end{equation}
where $E(\Omega,z)=\sqrt{\Omega_{\Lambda}+\Omega_{m}(1+z)^3}$.
Combining the expressions above, one obtains:
\begin{equation}
\Omega_{gw}(\nu_o)= \frac{8 \pi G}{3 c^2 H_0^3} \lambda \nu_o
\int^{z_{\sup}}_0 \frac{R_*(z)}{(1+z)E(\Omega,z)}
\frac{dE_{gw}(\nu_o(1+z))}{d\nu}dz
\end{equation}
or replacing the constants $G$ (gravitational constant) and $c$ (speed of light) by their usual values:
\begin{equation}
\Omega_{gw}(\nu_o)= 5.7 \times 10^{-56}(\frac{0.7}{h_0})^2 \lambda \nu_o
\int^{z_{\sup}}_0 \frac{R_*(z)}{(1+z)E(\Omega,z)}
\frac{dE_{gw}(\nu_o(1+z))}{d\nu}dz
\label{eq-omega}
\end{equation}
where $R_*$ is given for $h_0=0.7$.
The upper limit of the integral, which depends on both the maximal emission
frequency in the source frame and the maximal redshift of the
model of star formation history (usually $z_{\max}=5-6$), is given by:
\begin{equation}
z_{\sup}=
\left\lbrace
\begin{array}{ll}
z_{\max}    &   \hbox{  if } \nu_o < \frac{\nu_{\max} }{(1+z_{\max})}\\
\frac{\nu_{\max}}{\nu_o}-1 &   \hbox{  otherwise }\\
\end{array}
\right.
\label{eq-zsup}
\end{equation}
Consequently, the shape of the spectrum of any astrophysical background is
characterized by a cutoff at the maximal emission frequency and a
maximum at a frequency which depends on the shape of both the SFR and
the spectral energy density.

\section{The detection regimes}

Besides the spectral properties, it is important to study the nature of
the background. In the case of burst sources the integrated signal
received at $z=0$ from sources up to a redshift $z$, would show very
different statistical behavior depending on whether the duty cycle:

\begin{equation}
D(z)=\int^z_0 \bar{\tau} (1+z') \frac{dR}{dz'}(z') dz'
\label{eq-DC}
\end{equation}
defined as the ratio, in the observer frame, of the typical duration
of a single event $\bar{\tau}$, to the average time interval between successive events, is smaller or larger than unity.
When the number of sources is large enough for the time interval
between events to be small compared to the duration of a single event
($D>> 1$), the waveforms overlap to produce a continuous
background. Due to the central limit theorem, such backgrounds obey
the Gaussian statistic and are completely determined by their spectral
properties. They could be detected by data analysis
methods in the frequency domain such as the cross correlation
statistic presented in the next section \cite{all99}.
On the other hand, when the number of
sources is small enough for the time interval between events to be
long compared to the duration of a single event ($D << 1$), the
sources are resolved and may be detected by data
analysis techniques in the time domain (or the time frequency domain) such as
match filtering \cite{arn99,pra01}.
An interesting intermediate case arises when the time
interval between events is of the same order of the duration of a
single event. These signals, which sound like crackling popcorn, are
known as "popcorn noise". The waveforms may overlap but the statistic
is not Gaussian  anymore so that the amplitude on the detector at a
given time is unpredictable. For such signals, data analysis
strategies remain to be investigated \cite{dra03,cow05}, since the time
dependence is important and data analysis techniques in the frequency
domain, such as the cross correlation statistic, are not adapted.

\section{Models of astrophysical stochastic backgrounds}

In this section, we investigate two processes able to produce a continuous stochastic backgrounds in the frequency band of terrestrial interferometers.

\subsection{magnetars}

Rotating neutron stars (NSs) with a triaxial shape may have a time varying
quadrupole moment and hence radiate GWs at twice the rotational
frequency.
The total spectral gravitational energy emitted by a NS born with a rotational period $P_0$, and
which decelerates through magnetic dipole torques and GW emission, is given by:
\begin{equation}
\frac{dE_{gw}}{d \nu}=K \nu^3 (1+\frac{K}{\pi^2 I_{zz}} \nu^2)^{-1} \,\ \mathrm{with}\,\ \nu \in \lbrack 0-2/P_0 \rbrack
\label{eq-Enj_pulsar}
\end{equation}
where
\begin{equation}
K=\frac{192 \pi^4 G I^3}{5 c^2 R^6} \frac{\varepsilon^2}{B^2 \sin^2 \alpha}
\label{eq-K_pulsar}
\end{equation}
In this expression $R$ is the radius of the star, $\varepsilon=(I_{xx}-I_{yy})/I_{zz}$ the ellipticity, $I_{ij}$ the principal moment of inertia , $B$ the magnetic field and $\alpha$ the angle
between the rotation and the dipole axis.

In the original scenario of \cite{dun92,tho93}, super-strong crustal magnetic fields ($B \simeq 10^{14} - 10^{16}$ G) can be formed by dynamo action in proto neutron stars with very small rotational periods, larger than the break up limit around $0.5-1$ ms, but smaller than the convective overturn at 3 ms.
For these highly magnetized neutron stars, the distortion induced by the magnetic torque,
becomes significant, overwhelming the deformation due to the fast rotation.
In the case when the internal magnetic field is purely poloidal and matches to the dipolar field $B$ in the exterior, is given by \cite{bon96}:
\begin{equation}
\varepsilon_B=\beta \frac{R^8B^2 \sin^2 \alpha}{4GI_{zz}^2}
\label{eq-epsB_pol}
\end{equation}
According to the numerical simulations of \cite{bon96}, the distortion
parameter $\beta$, which depends on both the equation of state (EOS) and
the magnetic field geometry, can range between $1-10$ for a
non-superconducting interior to $100-1000$ for a type I superconductor
and even take values larger than $1000-10000$ for a type II
superconductor with counter rotating electric currents.
Assuming $R=10$ km, $I_{zz}=10^{45}$ g cm$^2$,  $P_0=1$ ms, $\beta=100$ and the average value given by the observation of SGR and AXP $B=10^{15}$ G ,  we obtain $\varepsilon_B \sim 2 \times 10^{-4}$ and $K \sim 3 \times 10^{37}$ erg Hz$^{-3}$.
In this case, the GW emission becomes negligible compared to the
magnetic torque and eq.~\ref{eq-Enj_pulsar} simplifies to
\begin{equation}
\frac{dE_{gw}}{d \nu} \sim K \nu^3
\label{eq-Enj_pulsar2}
\end{equation}
Considering that magnetars represent about $10\%$ of newborn NS, in
agreement with the estimates of \cite{kou98} and population synthesis of \cite{reg01a}, we obtain for the mass fraction of the  progenitors $\lambda= 0.1 \lambda_{NS}$ where the mass fraction of neutron star progenitors is given by:
\begin{equation}
\lambda_{NS}=\int_{8M_{\odot}}^{40M_{\odot}} {\xi(m)dm}=9 \times
10^{-3} {\mathrm  M}_{\odot}^{-1}
\label{eq-lambdaNS}
\end{equation}
where $\xi(m)$ is the modified A Salpeter IMF from eq.~\ref{eq-imf} and where we have assumed that NS progenitors have masses larger than 8 M$_ \odot$ \cite{hop06} and that stars with masses
larger than $40$ M$_ \odot$ give rise to black holes.
Evolution of such massive stars being
very fast, we can replace the previous expressions in
eq.~\ref{eq-omega}, and obtain:
\begin{equation}
\Omega_{gw}(\nu_o) \sim 4 \times 10^{-21} \nu_o^4 \int_0^{z_{\sup}(\nu_o)} dz
\frac{R_*(z)(1 + z)^2}{E(z)} 
\end{equation}
The energy density increases as $\nu_o^4$ at low frequencies and reaches a maximum of $\Omega_{gw} \sim 1 \times 10^{-10}$ around 1100 Hz.

\begin{figure}
\centering
\includegraphics[angle=0,width=0.8\columnwidth]{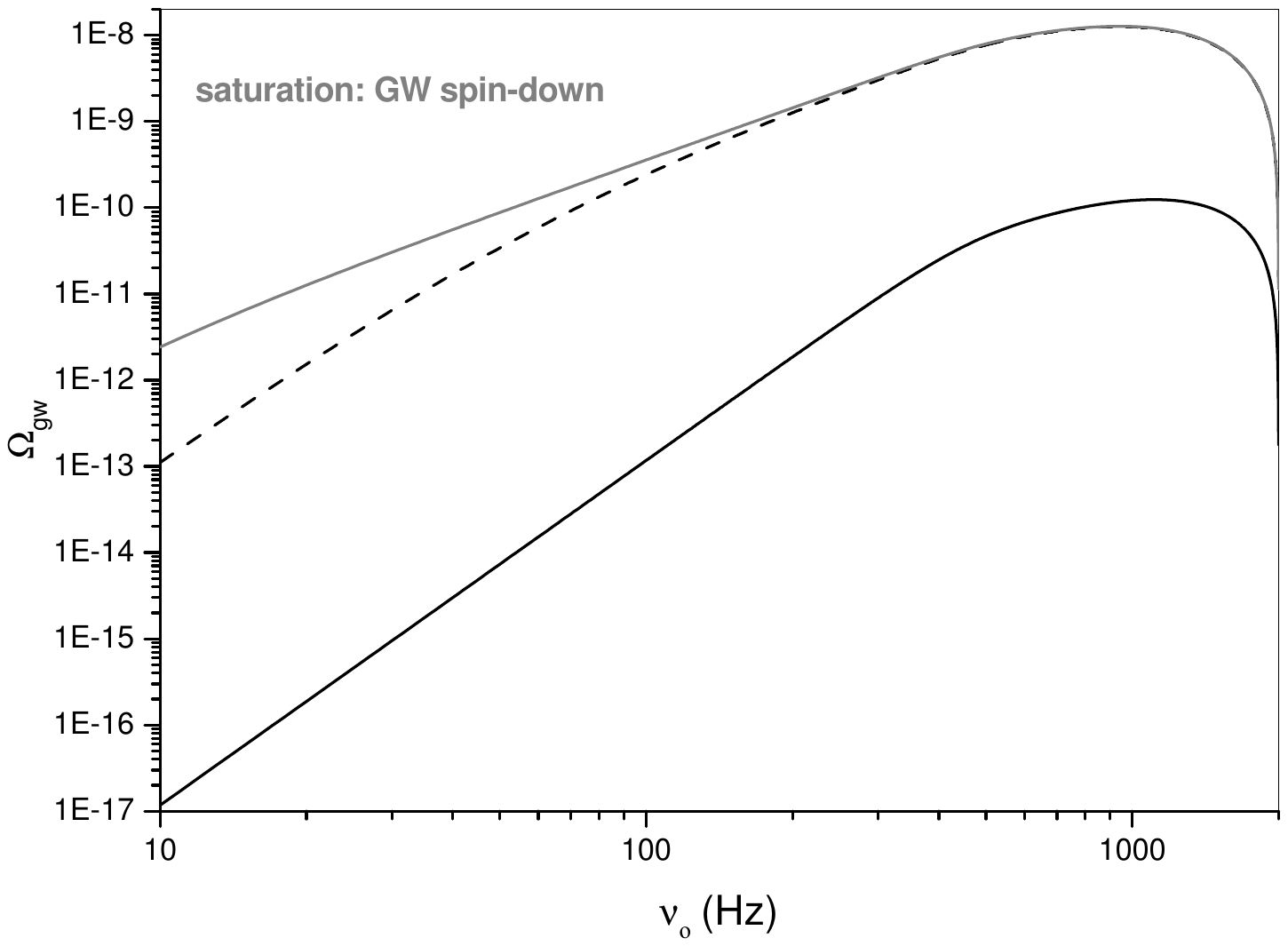}
\caption{energy density of the background produced by tri-axial
  rotating magnetars, as a function of the observed frequency,
  for the canonical model with a poloidal internal magnetic field and
  ms initial rotational period. The black continuous line corresponds to a model with $B=10^{15}$ G and $g=100$ (magnetic spindown regime), the grey continuous line to the purely gravitational spindown regime and the black dashed line to an intermediate case with  $B=10^{16}$ G and $g=1000$.
\label{fig-magnetar_pol_omega}}
\end{figure}

When the product $\beta B$ is large enough ($K>>\pi^2 I \nu^{-2}$),
GW emission becomes the most important process. In the saturation regime where the spindown is purely gravitational, eq.~\ref{eq-Enj_pulsar} simplifies to:
\begin{equation}
\frac{dE_{gw}}{d \nu} \sim \pi^{2} I \nu
\label{eq-Enj_pulsar3}
\end{equation}
and  the energy density increases as $\nu_o^2$ at low frequencies and reaches a maximum of $\Omega_{gw}
\sim 1.3 \times 10^{-8}$ around 1600 Hz (Fig.~\ref{fig-magnetar_pol_omega}).
It has been suggested that the internal magnetic field
could be dominated by the toroidal component \cite{cut02,ste05}.
In this case
\begin{equation}
\varepsilon_B= 1.6 \times 10^{-4} <B_{t,16}^2>
\label{eq-epsB_tor}
\end{equation}
where $<B_{t,16}^2>$ is the mean value of the internal toroidal component in unit of $10^{16}$ G,
and whether the rotational energy is dissipated due to dipole or GW emission
depends on the ratio $B_t^2/B$.

\subsection{Double neutron star coalescence}

The merger of two neutron stars, two black holes or a neutron star and
a black hole are among the most important sources of gravitational
waves, due to the huge energy released in the process.
In particular, double neutron stars (DNSs) may radiate about $10^{53}$ erg in the
last seconds of their inspiral trajectory.
In a recent work, \cite{reg06b,reg07} used Monte Carlo
simulations to calculate the contribution of DNSs to the stochastic
background, in the frequency band of ground based interferometers,
which corresponds to the last $ \sim 1000$ s before the last stable
orbit, when more than 96\% of the gravitational energy is
released.
At that time, the system has been circularized through GW
emission and the spectral energy density is given in the quadrupolar approximation by:
\begin{equation}
\frac{dE_{gw}}{d\nu} = K_b \nu^{-1/3} \,\ \mathrm{with}\,\ \nu \in \lbrack 10-\nu_{LSO} \rbrack
\end{equation}
where
\begin{equation}
K_b = \frac{(G \pi)^{2/3}}{3} \frac{m_1m_2}{(m_1+m_2)^{1/3}}
\end{equation}
For double neutron stars with masses $m_1 = m_2 =
1.4$ M$_\odot$, one obtains $K_b=5.2 \times 10^{50}$ erg
Hz$^{-2/3}$ and the gravitational frequency at the last stable orbit is
assumed to be $\nu_{LSO}=1.57$ kHz \cite{sat00}.

\begin{figure}
\centering
\includegraphics[angle=0,width=0.8\columnwidth]{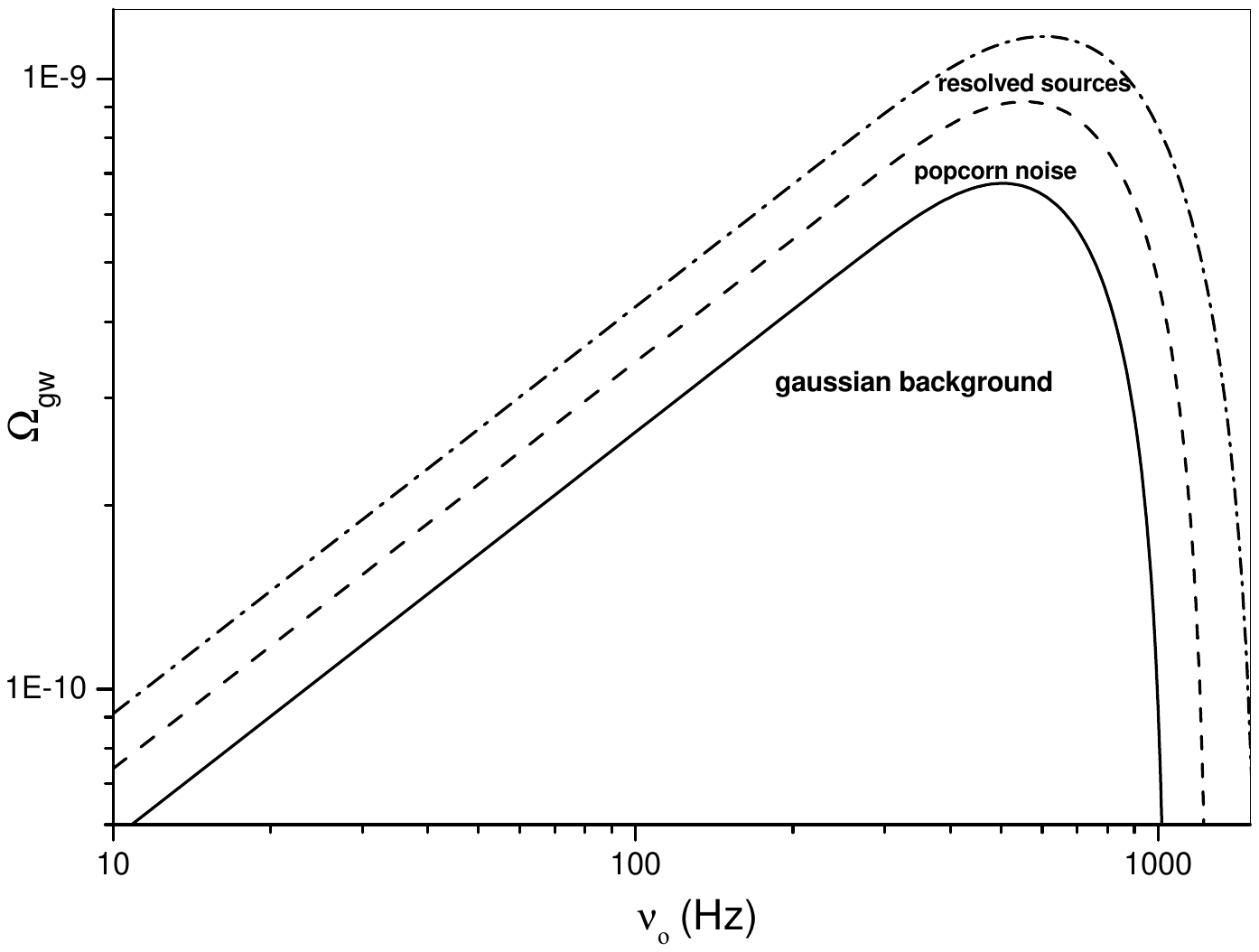}
\caption{energy density of the continuous background produced by DNS
  coalescences at $z > 0.5$ (continuous line) and of the popcorn
  contribution corresponding to sources between $z=0.25-0.5$ (dashed
  line). The signal from the whole population down to $z=0$ is also plotted for
  comparison (dot-dashed line).
\label{fig-DNS_omega}}
\end{figure}

The merging occurring long after the formation of the system of massive stars,
the coalescence rate per unit of comoving volume,
which will replace $R_*(z)$ in eq.~\ref{eq-rate} and eq.~\ref{eq-omega} is given by:
\begin{equation}
R_c(z)=\int\frac{1+z}{1+z_f} R_*(t_c(z)-t_d)P(t_d)dt_d
\end{equation}
where $z_f$ is the redshift of formation of the progenitors and $t_c(z)$ the cosmic time at the redshift of coalescence $z$.
The probability distribution of the delay time is usually parameterized by \cite{dfp06,and05,sha07}:
\begin{equation}
P_d(t_d) \propto \frac{1}{t_d} \,\ \mathrm{with} \,\ t_d > \tau_0
\end{equation}
where the minimal delay time $\tau_0 \simeq 20 Myr$ corresponds roughly to the time it takes
for massive binaries to evolve into two neutron stars \cite{bel06}.
The mass fraction converted into the progenitors is given by the product
$\lambda_{b} = \beta_{NS} f_b \lambda_{NS}$, where $\beta_{NS}$ is the
fraction of binaries which remains bounded after the second supernova
event, $f_b$ the fraction of massive binaries formed among all stars
and $\lambda_{NS}$ the mass fraction of NS progenitors.
In our reference model we assume $\beta_{NS}f_b= 0.03$ \cite{dfp06}.
The merging rate in our galaxy at time $t$ is given by:
\begin{equation}
r_{\mathrm{mw}}(t)=\lambda_b \int^t_{\tau_0} \rho_{\mathrm{mw}}(t-t_d) P_d(t_d) dt_d
\end{equation}
where $\rho_{mw}(t) = n_{\mathrm{mw}}^{-1} R_{mw}(t)$ is the galactic star formation rate in M$_\odot$ yr$^{-1}$.
Using the model of \cite{nag06} for the star formation history in the disk, namely
\begin{equation}
R_{mw}(t)=0.056 \frac{t}{4.5} e^{-\frac{t}{4.5}} \,\ \mathrm{(M}_\odot \mathrm{yr}^{-1}\mathrm{Mpc}^{-3})
\end{equation}
where $t$ is in Gyr, and assuming a density of milky-way equivalent galaxies of $n_{\mathrm{mw}}=0.01$ Mpc$^{-3}$,
we find an actual rate of $r_{\mathrm{mw}}^o=r_{\mathrm{mw}}(13.5) \sim 3\times 10^{-5}$ yr$^{-1}$,
in agreement with the expectations derived from statistical studies and population synthesis \cite{kim04,dfp06,bel06a}.
We find that sources at redshifts $z>0.5$ contribute to a truly continuous stochastic background, while sources at
redshifts $0.25<z<0.5$  are responsible for a popcorn noise, with duty
cycles of 1 and 0.1 respectively.
The energy density
\begin{equation}
\Omega_{gw} (\nu_0)= 8.6 \times 10^{-10} \nu_o^{2/3}
\int^{z_{\sup}}_0 \frac{R_c(z)}{(1+z)^{4/3}E(z)} dz
\label{eq-omega_DNS}
\end{equation}
reaches a maximum of $\Omega_{gw} \sim 7 \times  10^{-10}$  around  500 Hz for
the continuous contribution and of $\Omega_{gw} \sim 9 \times
10^{-10}$ around 550 Hz for the popcorn background (Fig.~\ref{fig-DNS_omega}).
The total background, including the nearest sources down to $z \sim 0$
is slightly higher, with a maximum of $\Omega_{gw} \sim 1.2 \times
10^{-9}$ at 600 Hz.

\section{Constraints with the next generations of detectors}

We now discuss the constraints one can expect to put on some simple models,
with the next generations of terrestrial detectors.

\subsection{detection}
The optimal strategy to search for a stochastic
background, which can be confounded with the intrinsic noise
background of the instrument, is to cross correlate measurements of multiple
detectors.
The optimized $S/N$ ratio for an integration time $T$ is given by \cite{all99}:
\begin{equation}
(\frac{S}{N})^2 = \frac{9 H_0^4}{50 \pi^4} T \int_0^\infty df \frac{\gamma^2(f)\Omega_{\mathrm{gw}}^2(f)}{f^6 P_1(f)P_2(f)}
\end{equation}
where $P_1(f)$ and $P_2(f)$ are the power spectral noise
densities of the two detectors and $\gamma$ is the normalized overlap
reduction function, characterizing the loss of sensitivity due to
the separation and the relative orientation of the detectors.
We use the advanced LIGO sensitivity as an example of the second generation,
the Einstein Telescope as an example of the third generation,
and the LIGO Hanford/Livingston pair (H1-L1) as an example of separated detectors.
For separated detectors the signal to noise ratio ($S/N$) is
calculated between $10-150$ Hz and for co-aligned and coincident detectors,
between $10-500$ Hz.
We assume an observation time of $T=3$ yr and detection threshold of $S/N=1$ and $S/N=5$.

\subsection{magnetars}
\begin{figure}
\centering
\includegraphics[angle=0,width=0.8\columnwidth]{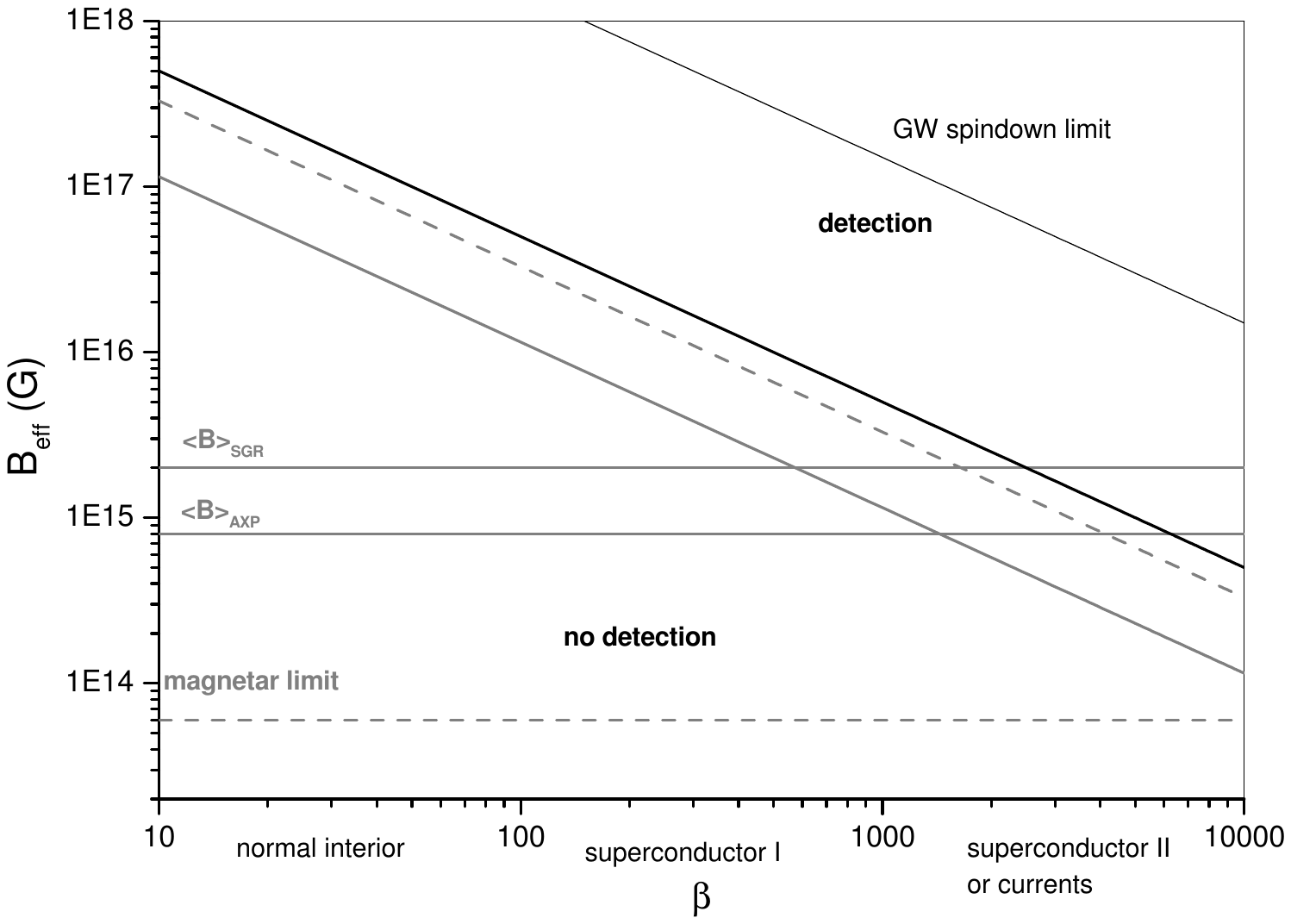}
\caption{magnetars: constraints on the dipole external magnetic field $B_{\mathrm{eff}}$
and the distortion parameter $\beta$ with co-aligned and coincident detectors and integration time $T=3$ yr. The black continuous line corresponds to advanced detectors and a detection threshold $S/N=1$, the light grey lines to 3rd generation detectors and $S/N=1$ (continuous) and  $S/N=5$ (dashed).
The grey continuous line shows the limit above which the spindown is almost purely gravitational and corresponds to $S/N \sim 2$ with advanced detectors and $S/N \sim 23$ with the third generation.
The average values of $B_{\mathrm{eff}}$ for observed AXPs and SGRs as well as the magnetar limit,
are also plotted for comparison.

\label{fig-magnetar_constraints}}
\end{figure}

We investigate the constraints
on the mean value of the effective external magnetic field $B_{eff}$
and the deformation parameter $\beta$, for the \emph{canonical model} with ms initial periods (dynamo process).
Fig.~\ref{fig-magnetar_constraints} shows the detection limit in the
plane $B_{eff}-\beta$ for advanced and third generation co-aligned and coincident detectors and for thresholds $S/N=1$ and 5.
At least with the third generation, we should be able to exclude the highest possible values
of the magnetic field ($B_\mathrm{eff}>10^{16} G$) and models with type II
superconducting interior or with counter rotating electric currents.
We could also put constraints on the internal toroidal component of the magnetic field $B_t$.
If there is no detection, we should be able to confirm or rule out models for which the spindown is dominated by GW emission, for instance with $B_t>>B$ \cite{ste05}, which give $S/N \sim 2$ with advanced detectors and $S/N \sim 23$ with the third generation.
With the first generation of detectors, unfortunately, it won't be possible to say anything, since we are well below the detection threshold, even for the most optimistic case when the spindown of the star is due to GW emission only.

\subsection{Double neutron star coalescence}
We use the model of section 4 to investigate the constraints on the fractions $f_b$ and $\beta_{ns}$,
or equivalently on the galactic coalescence rate $R_{mw}$ (Table~\ref{table-DNS_constraints}), given by statistical studies and source modeling in the range $10^{-4}- 10^{-6}$ yr$^{-1}$, more likely around $10^{-5}$ yr$^{-1}$ \cite{kim04,dfp06,bel06a}.
With advanced coaligned and coincident detectors, we should be able  to exclude the most optimistic predictions. It would become extremely interesting with the third generation,
since we expect to reach the actual theoretical expectations with separated detectors and get very close to the lower limit with co-aligned and coincident detectors.
If there is no detection with the third generation, we may have to review our models of binary evolution.

\begin{table}
\centering
\begin{tabular}{lccc}
\noalign{\smallskip}
\hline
\noalign{\smallskip}
& $(S/N)^{-1} f_b \beta_{ns}$  & $(S/N)^{-1} R_{mw}$ yr$^{-1}$ & $R_{mw}$ yr$^{-1} (S/N=5)$\\
\noalign{\smallskip}
\hline
\noalign{\smallskip}
inital (HL) & 138.3 & 1.3 & 6.6 \\
initial (c-co) & 1.55 & 0.015 & 0.074 \\ 
Ad (HL) & 0.047 & $4.5 \times 10^{-4}$ & 0.0022\\
Ad (c-co) & 0.0025 &  $2.4 \times 10^{-5}$ & $1.2 \times 10^{-4}$\\
3rd (HL) & $4.5 \times 10^{-4}$ & $4.3 \times 10^{-6}$ & $2.2 \times 10^{-5}$ \\
3rd (c-co) & $1.6 \times 10^{-4}$ & $1.5 \times 10^{-6}$ & $7.5 \times 10^{-6}$\\
\noalign{\smallskip}
\hline
\end{tabular}
\caption{Constraints on the product $f_b \beta_{ns}$ and on the galactic coalescence rate $R_{mw}$ of double neutron stars for different generation of detectors, for an integration time $T=3$ yr and detection thresholds $S/N=1$ (column 1 and 2) and $S/N=5$ (third column). HL indicates a pair of detectors separated such as the LIGO Hanford-Livingston pair, and c-co a pair of co-aligned and coincident detectors. One can obtain the constraints for any detection threshold from  columns 1 and 2  by multiplying the values given in column 1 and 2 ($S/N=1$) by the $S/N$. Column 3 corresponds to $S/N=5$)}
\label{table-DNS_constraints}
\end{table}

\section{Conclusions}
In this article, we reviewed the spectral and statistical properties of astrophysical backgrounds
and presented two promising models for ground based detectors, as well as the constraints
one can expect to put on the source parameters with the next generation of detectors.
We showed that the energy density of astrophysical backgrounds is characterized
by an increase at low frequencies (usually a power law), a maximum and a cutoff,
at the maximum emission frequency in the source frame.
For most of our models the peak occurs at kHz frequencies,
where the sensitivity of pairs of separated detectors drops
significantly.
The best strategy may be to use co-aligned and coincident detectors
such as the two LIGO Hanford interferometers;
the issue is the presence of correlated noise, but new techniques are under
development in the LIGO collaboration which should allow to use this pair in the near future \cite{fot08}.
On the other hand, the stochastic background may result in a popcorn
noise or may be anisotropic at close redshifts, but again adequate detection
strategies are also under investigation \cite{bal06,mit08}.
We showed that we could already obtain
interesting astrophysical results on simple models with advanced and third generation
detectors, such as upper limits on the
magnetic field or the ellipticity of magnetars or
the coalescence rate of compact binaries.
On the other hand we expect to be able to rule out or confirm some extreme models
such as  pure GW spindown in magnetars.
Work is currently in progress to investigate a larger range of models and parameters and it will be reported in a future paper.
What is particularly interesting with stochastic backgrounds, is that
we can put constraints on the mean value of the population,
and not just on some particular sources that may be in the tail of the probability distributions
of the ellipticity or the magnetic field.

\end{document}